\documentclass{ws-ijmpd_new}
\begin{document}

%

\def\nocropmarks{\vskip5pt\phantom{cropmarks}}

 \let\trimmarks\nocropmarks      

%


\def\gtap{\mathrel{ \rlap{\raise 0.511ex \hbox{$>$}}{\lower 0.511ex
   \hbox{$\sim$}}}} \def\ltap{\mathrel{ \rlap{\raise 0.511ex
   \hbox{$<$}}{\lower 0.511ex \hbox{$\sim$}}}} 
\newcommand{\beq}{\begin{equation}}
\newcommand{\dd}{\partial}
\newcommand{\eeq}{\end{equation}}
\newcommand{\bea}{\begin{eqnarray}}
\newcommand{\eea}{\end{eqnarray}}
\newcommand{\lsim}{\stackrel{<}{\scriptstyle \sim}}
\newcommand{\gsim}{\stackrel{>}{\scriptstyle \sim}}
\newcommand{\La}{{\cal L} }
\newcommand{\half}{\frac{1}{2} }
\newcommand{\bra}[1]{\langle #1 |}
\newcommand{\ket}[1]{| #1 \rangle}
\newcommand{\sprod}[2]{\langle #1 , #2 \rangle}
\newcommand{\braket}[2]{\langle #1 | #2 \rangle}
\newcommand{\eq}[1]{eq.(\ref{#1})}
\newcommand{\dpar}[2]{\frac{\partial #1}{\partial #2}}
\newcommand{\vpar}[2]{\frac{\delta #1}{\delta #2}}
\newcommand{\ddpar}[2]{\frac{\partial^2 #1}{\partial #2^2}}
\newcommand{\vvpar}[2]{\frac{\delta^2 #1}{\delta #2^2}}
\newcommand{\eV}{\mbox{$ \ \mathrm{eV}$}}
\newcommand{\KeV}{\mbox{$ \ \mathrm{KeV}$}}
\newcommand{\MeV}{\mbox{$ \ \mathrm{MeV}$}}
\newcommand{\probm}{\mbox{$ \ \langle P_m \rangle$}}

\def\s{\mbox{\boldmath$\displaystyle\mathbf{\sigma}$}}
\def\J{\mbox{\boldmath$\displaystyle\mathbf{J}$}}
\def\K{\mbox{\boldmath$\displaystyle\mathbf{K}$}}

\def\beq{\begin{eqnarray}}
\def\eeq{\end{eqnarray}}

\def\cdash{$^{\raisebox{-0.5pt}{\hbox{--}}}$}         

%


\markboth{Alexander Kusenko}{Pulsar kicks from neutrino oscillations}

%
\catchline{}{}{}
%

\title{Pulsar kicks from neutrino oscillations} 


\author{\footnotesize Alexander Kusenko} 

\address{Department of Physics and Astronomy, \\ 
         University of California, Los
          Angeles, CA 90095-1547, USA} 


\maketitle

\begin{abstract}

Neutrino oscillations in a core-collapse supernova may be responsible for
the observed rapid motions of pulsars.  Given the present bounds on the
neutrino masses, the pulsar kicks require a sterile neutrino with mass
2--20~keV and a small mixing with the active neutrinos.  The same particle
can be the cosmological dark matter.  Its existence can be confirmed the by
the X-ray telescopes if they detect a 1--10~keV photon line from the decays
of the relic sterile neutrinos.  In addition, one may be able to detect
gravity waves from a pulsar being accelerated by neutrinos in the event of
a nearby supernova.

\end{abstract}


\section{Introduction}

There is an intriguing possible connection between two long-standing
astrophysical puzzles: the origin of pulsar velocities and the nature of
cosmological dark matter.  The evidence for dark matter is extremely
strong; its existence requires at least one new particle that is not a part
of the Standard Model.  If this new particle is a singlet fermion that has
a small mixing with neutrinos, its emission from a supernova could be
anisotropic.\cite{ks97,fkmp} The anisotropy could explain the observed
pulsar velocities.  The purpose of this review is to explore this
explanation of the pulsar kicks.

\subsection{Pulsar velocities}

The space velocities of pulsars are measured either by observation of their
angular proper motions,\cite{astro1} or by measuring the velocity of an
interstellar scintillation pattern as it sweeps across the
Earth.\cite{astro2,astro2_1}  Each of the two methods has certain
advantages.  Using the former method, one can get very precise measurements
with the help of a high-resolution radio interferometer, but such
observations take a long time.  Measuring the velocity of a scintillation
pattern can be done quickly, but the inference of the actual pulsar
velocity must rely on some assumptions about he distribution of scattering
material along the line of sight. (For instance, if the density of
scatterers is higher near Earth, the pattern speed is less than the pulsars
speed.)  In addition to observational errors, one has to take into account
various selection effects.  For example, fast and faint pulsars are
under-represented in the data as compared with the slow and bright ones.
Therefore, one has to carefully model the pulsar population to calculate
the three-dimensional distribution of pulsar velocities corresponding to
the observed two-dimensional projection of their proper
motion.\cite{astro_3d,astro_15}

Based on the data and population models, the average velocity estimates
range from 250~km/s to 500~km/s.\cite{astro1}\cdash\cite{astro_15} The
distribution of velocities is non-gaussian, and there is a substantial
population of pulsars with velocities in excess of 700~km/s.  Some 15\% of
pulsars\cite{astro_15} appear to have velocities greater than 1000~km/s,
while the fastest pulsars have speeds as high as 1600~km/s.  Obviously, an
acceptable mechanism for the pulsar kicks must be able to explain these
very fast moving pulsars.

Pulsars are born in supernova explosions, so it would be natural to look
for an explanation in the dynamics of the
supernova\cite{explosion}.  However, state-of-the-art \mbox{3-dimensional}
numerical calculations\cite{fryer} show that even the most extreme
asymmetric explosions do not produce pulsar velocities greater than
200~km/s.  Earlier 2-dimensional calculations\cite{sn_2D} claimed a
somewhat higher maximal pulsar velocity, up to 500~km/s.  Of course, even
that was way too small to explain the population of
pulsars with speeds (1000--1600)~km/s.  Recent three-dimensional
calculations by Fryer\cite{fryer} show an even stronger discrepancy than
the earlier numerical calculations of the supernova.  

The hydrodynamic kick could be stronger if some large initial asymmetries
developed in the cores of supernova progenitor stars prior to their
collapse.  Goldreich {\em et~al.}\cite{gls} have suggested that unstable
g-modes trapped in the iron core by the convective burning layers and
excited by the $\epsilon$-mechanism may provide the requisite asymmetries.
However, according to recent numerical calculations,\cite{mbh} the
$\epsilon$-mechanism may not have enough time to significantly amplify the
g-modes prior to the collapse.  A different kind of the seed unisotropies
may develop from the north-south asymmetry in the neutrino heating due to a
strong magnetic field.\cite{sato}  If these asymmetries grow sufficiently
during the later phases of the supernova, they may be relevant for the
pulsar kicks.

Evolution of close binaries\cite{b} and asymmetric emission of radio
waves\cite{ht} have been considered as possible causes of the rapid pulsar
motions.  However, both of these explanations fail to produce a large
enough effect.

Most of the supernova energy, as much as 99\% of the total $10^{53}$~erg
are emitted in neutrinos.  A few per cent anisotropy in the distribution of
these neutrinos would be sufficient to explain the pulsar kicks.
Alternatively, one needs (and, one is apparently lacking) a much larger
asymmetry in what remains after the neutrinos are subtracted from the
energy balance.  The numerical calculations of the supernova assume that
neutrino distribution is isotropic.  What if this is not true?

Since the total energy released in supernova neutrinos is $E\sim 3\times
10^{53}$erg, the outgoing neutrinos carry  the total momentum 
\beq
p_{\nu,{\rm total}} 
\sim 1\times 10^{43} {\rm g\,cm/s}. 
\eeq
A neutron star with mass 1.4$M_\odot$ and $v= 1000$~km/s has momentum
\bea
p_* =  (1.4 M_\odot ) v  & \approx & 3 \times 10^{41} \left (
\frac{v}{1000\,   {\rm km/s}} \right ) {\rm g\,cm/s} \nonumber \\
& \approx & 0.03 \left ( \frac{v}{1000\, {\rm km/s}} \right ) 
p_{\nu,{\rm total}} . 
\eea
A few per cent asymmetry in the neutrino distribution is, therefore,
sufficient to explain the observed pulsar velocities.  What could cause
such an asymmetry?  The obvious suspect is the magnetic field, which can
break the spherical symmetry and which is known to have an effect on weak
interactions.  We will examine this possibility in detail.

\subsection{Neutrinos in the Standard Model and beyond} 

The number of light ``active'' left-handed neutrinos -- three -- is well
established from the LEP measurements of the Z-boson decay width.  In the
Standard Model, the three active neutrinos fit into the three generations
of fermions. In its original form the Standard Model described massless
neutrinos.  The relatively recent but long-anticipated discovery of the
neutrino masses has made a strong case for considering right-handed
neutrinos, which are SU(3)$\times$SU(2)$\times$U(1) singlets.  The number
of right-handed neutrinos may vary and need not equal to
three.\cite{kayser,2right-handed} Depending on the structure of the
neutrino mass matrix, one can end up with none, one, or several states that
are light and (mostly) sterile, {\em i.e.}, they interact only through
their small mixing with the active neutrinos.

Unless some neutrino experiments are wrong, the present data on neutrino
oscillations cannot be explained without sterile neutrinos.  Neutrino
oscillations experiments measure the differences between the squares of
neutrino masses, and the results are:\cite{kayser} one mass squared
difference is of the order of $10^{-5}$(eV$^2$), the other one is
$10^{-3}$(eV$^2$), and the third is about $1\,$(eV$^2$).  Obviously, one
needs more than three masses to get the three different mass splittings
which do not add up to zero.  Since we know that there are only three
active neutrinos, the fourth neutrino must be sterile.  However, if the
light sterile neutrinos exist, there is no compelling reason why their
number should be limited to one.  We will see that a sterile neutrino
required to explain the pulsar kicks and dark matter simultaneously must
have a mass in the 2--20~keV range and a very small mixing.
 
In addition to explaining the neutrino oscillation data, pulsar kicks, and
dark matter, theoretical models have invoked sterile neutrinos for various
other reasons.  For example, Farzan {\em et al.}\cite{farzan} used eV--keV
sterile neutrinos to produce the mass matrices with certain
properties, such as nearly bimaximal mixing\cite{bimaximal} of active
neutrinos.  Sterile neutrinos with mass around 200 MeV could reionize the
universe even before the star formation,\cite{reionization} as early as the
WMAP data suggest.\cite{wmap}

In general, the SU(3)$\times$SU(2)$\times$U(1) singlet (sterile) neutrino
is not an eigenstate of the mass matrix.  The mass eigenstates are linear
combinations of the weak eigenstates.  Let us assume, for example, that the
singlet neutrino has a non-zero mixing with the electron neutrino, but that
the other mixing angles are zero or very small.  Then one finds that the mass
eigenstates have a simple expression in terms of the weak eigenstates: 
\begin{eqnarray}
| \nu_1 \rangle & = & \cos \theta_m \, | \nu_e \rangle - \sin \theta_m \, |
\nu_s  \rangle \\ 
| \nu_2 \rangle & = & \sin \theta_m \, | \nu_e \rangle + \cos \theta_m \, |
\nu_s \rangle  . 
\label{eigenstates}
\end{eqnarray}

If the mixing angle $\theta_m $ is small, one of the mass eigenstates,
$\nu_1$ behaves very much like a pure $\nu_e$, while the other, $\nu_2$, is
practically ``sterile'', which means it has weak interactions suppressed by
a factor $(\sin^2 \theta_m)$ in the cross section.

As discussed below, the sterile neutrinos that can kick the pulsars should
have mass in the 2--20~keV range, and they should also have a small mixing
$(\sin \theta \sim 10^{-4})$ with ordinary neutrinos, for example, the
electron neutrino.  Theoretical models of neutrino masses can readily
produce a sterile neutrino with the required mass and
mixing.\cite{farzan,babu}

\subsection{Why a sterile neutrino can give the pulsar a kick}

\label{sec_why}

Given the lack of a ``standard'' explanation for the pulsar kicks, one is
compelled to consider alternatives, possibly involving new physics.  One
reason why the standard explanation fails is because most of the energy is
carried away by neutrinos, which escape isotropically.  The remaining
momentum must be distributed with a substantial asymmetry to account for
the large pulsar kick.  In contrast, we saw that only a few per cent
anisotropy in the distribution of neutrinos would give the pulsar a kick
of required magnitude.

Neutrinos are always {\em \sf produced } with an asymmetry, but they
usually {\em \sf escape} isotropically.  The asymmetry in production comes
from the asymmetry in the basic weak interactions in the presence of a
strong magnetic field.  Indeed, if the electrons and other fermions are
polarized by the magnetic field, the cross section of the urca processes,
such as $n+e^+ \rightleftharpoons p+ \bar \nu_e$ and
$p+e^-\rightleftharpoons n+ \nu_e $, depends on the orientation of the
neutrino momentum:
\beq 
\sigma ( \uparrow e^-, \uparrow \nu ) \neq  \sigma ( \uparrow e^-,
\downarrow \nu ) 
\label{sigma_up_down}
\eeq
Depending on the fraction of the electrons in the lowest Landau level, this
asymmetry can be as large as 30\%, which is, seemingly, more than one needs
to explain the pulsar kicks.\cite{drt}  However, this asymmetry is
completely washed out by scattering of neutrinos on their way out of the
star.\cite{eq}  This is intuitively clear because, as a result of
scatterings, the neutrino momentum is transferred to and shared by the
neutrons.  In the approximate thermal equilibrium, no asymmetry in the
production or scattering amplitudes can result in a macroscopic
momentum anisotropy.  This statement can be proved rigorously.\cite{eq} 

However, if the neutron star cooling produced a particle whose interactions
with nuclear matter were {\em \sf even weaker} than those of ordinary
neutrinos, such a particle could escape the star with an anisotropy equal
its production anisotropy.  The state $\nu_2$ in equation
(\ref{eigenstates}), whose interactions are suppressed by $(\sin^2
\theta_m)$ can play such a role.  It is intriguing that the same particle
can be the dark mater.

The simplest realization of this scenario is a model with only one singlet
fermion in which the mass eigenstates are admixtures of active and
sterile neutrinos, as in equation (\ref{eigenstates}).
For a sufficiently small mixing angle $\theta_m$ between $\nu_e$ and
$\nu_s$, only one of the two mass eigenstates, $\nu_1$, is trapped in the
core of a neutron star.  The orthogonal state, $\nu_2$, escapes from the
star freely.  This state is produced in the same basic urca reactions
($n+e^+ \rightleftharpoons p+ \bar \nu_e$ and
$p+e^-\rightleftharpoons n+ \nu_e $) 
with the effective Lagrangian coupling equal the weak coupling
times $\sin \theta_m$.  The production rate can be greatly enhanced if
active neutrinos undergo a resonant conversion into the sterile neutrinos
at some density.  This effect will play an important role in some range of
parameters, although this kind of enhancement is not necessary for the
pulsar kick.

We will consider two ranges of parameters, for which the $\nu_e \rightarrow
\nu_s$ oscillations occur on and off resonance. First, we will suppose that
a resonant oscillation occurs somewhere in the core of a neutron star.
Then the asymmetry in the neutrino emission comes from a shift in the
resonance point depending on the magnetic field.\cite{ks97}  The
temperature gradient is then responsible for the asymmetry in the momentum
carried by neutrinos.  Second, we will consider a resonance outside the
dense core, where the asymmetry has a somewhat different origin: it comes
from the uncompensated momentum deposition by active neutrinos on one side
of the star, while the corresponding neutrinos on the other side propagate
the same layer of matter as sterile neutrinos.  Finally, we will consider
the off-resonance case,\cite{fkmp} in which the asymmetry comes directly
from the weak processes, as in eq.~(\ref{sigma_up_down}).  However, before
discussing the emission of sterile neutrinos from a supernova, let us
briefly review their role in cosmology and the cosmological limits on their
masses and mixing angles.\cite{dolgov_nu}

\section{Relic sterile neutrinos as dark matter} 

Very few hints exist as to the nature of cosmological dark matter.  We know
that none of the Standard Model particles can be the dark matter, and we
also know that the dark matter particles should either be weakly
interacting or very heavy (or both).  Theoretical models have provided
plenty of candidates.  For example, supersymmetric
extensions of the Standard Model predict the existence of many new
particles, including two dark matter candidates: the lightest
supersymmetric particle (LSP) and the SUSY Q-balls.  These are plausible,
theoretically motivated dark-matter candidates, as are many
others.\cite{dm_review} 

However, if one seeks a {\em minimal } solution to the dark matter
problem, sterile neutrinos offer a unique possibility: one can add just one
dark-matter particle to the Standard Model, as long as it is an
SU(3)$\times$SU(2)$\times$U(1) singlet.  Gauge singlets can be produced
in weak interactions through their mixing with ordinary neutrinos.  

As discussed below, we are interested in masses in the 1--20 keV range and
 small mixing angles.  Such sterile neutrinos, which interact with
primordial plasma only via mixing, could never have been in thermal
equilibrium in the early universe.  They could be produced from active
neutrinos through oscillations.  However, at very high temperatures active
neutrinos have frequent interactions in plasma.  Matter and the 
quantum damping effects inhibit neutrino oscillations.\cite{stodolsky}
The mixing of sterile neutrinos with one of the active species in plasma 
can be represented by an effective, density and temperature dependent 
mixing angle\cite{dw,Fuller,dolgov_hansen}: 
\begin{equation}
\sin^2 2 \theta_m = 
\frac{(\Delta m^2 / 2p)^2 \sin^2 2 \theta}{(\Delta m^2 / 2p)^2 \sin^2 
2 \theta + ( \Delta m^2 / 2p \cos 2 \theta - V_m-V_{_T})^2}, 
\label{sin2theta}
\end{equation}
Here $V_m$ and $V_T$ are the effective matter and temperature potentials.
In the limit of small angles and small lepton asymmetry, the mixing angle
can be approximated as 
\beq
\sin 2 \theta_m \approx
\frac{\sin 2 \theta}{1+ 0.27 \zeta  \left( \frac{T}{100 \,
\rm MeV} \right)^6 \left( \frac{{\rm keV}^2}{\Delta m^2} \right )
}
\eeq
where $\zeta =1.0$ for mixing with the electron neutrino and  $\zeta 
=0.30$ for $\nu_\mu$ and $\nu_\tau$.  

Obviously, thermal effects suppress the mixing significantly for
temperatures $T> 150 \, (m/{\rm keV})^{1/3}\,$MeV.  Since the singlet
neutrinos interact only through mixing, all the interaction rates are
suppressed by the square of the mixing angle, $\sin^2 \theta_m $.  It is
easy to see that these sterile neutrinos are {\em never} in thermal
equilibrium in the early universe.  Thus, in contrast with the case of the
active neutrinos, the relic population of sterile neutrinos is not a result
of a freeze-out.  One immediate consequence of this observation is that the
Gershtein--Zeldovich bound\cite{gz} and the Lee--Weinberg bound\cite{lw} do
not apply to sterile neutrinos.

\begin{figure}[ht]
\centerline{\epsfxsize=4.9in\epsfbox{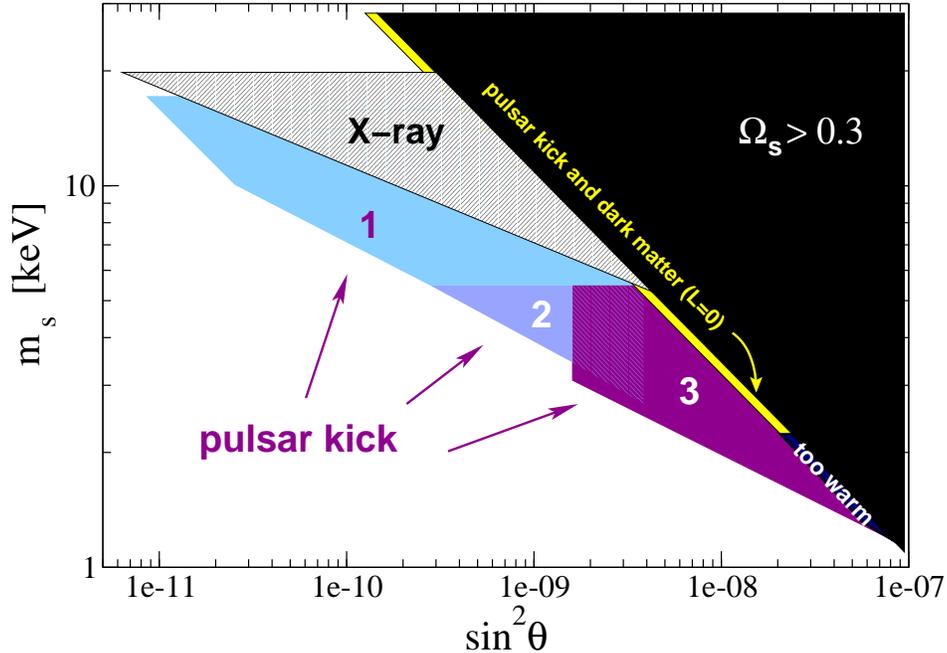}}   
\caption{The range of the sterile neutrino mass and mixing angle.  Regions
  1 and 2 correspond to parameters consistent with the pulsar kicks due to
  resonant MSW oscillations deep in the core (1) or outside the core (2).
  These two possibilities are discussed in sections \ref{sec_res_core} and
  \ref{sec_res_lessdense}, respectively.  Region 3 corresponds to
  off-resonance active--sterile conversions in the core (see section
  \ref{sec_offres}).  Cosmological bounds and the exclusion region due to
  X-ray observations are shown as well.  The cosmological bounds shown here
  assume that the reheat temperature after inflation was higher than 1~GeV
  and that the lepton asymmetry of the universe is small ({$L\ll
  10^{-3}$}).  }
\label{figure:range}
\end{figure}

Sterile neutrinos are produced through oscillations of active neutrinos.
The relation between their mass and the abundance is very different from
what one usually obtains in freeze-out.  One can trace the production of
sterile neutrinos in plasma by solving the Boltzmann equation for the
distribution function $f(p,t)$:
\bea
\left ( \frac{\dd}{\dd t}-H p \frac{\dd}{\dd p}\right ) f_s(p,t) & \equiv
& x H\dd_x f_s =  \\
& & \Gamma_{(\nu_a\rightarrow \nu_s)}
\left ( f_a(p,t) - f_s(p,t) \right ),  
\eea
where $H$ is the Hubble constant, $x=1\,$MeV$a(t)$, $a(t)$ is the scale
factor, and $\Gamma$ is the probability of conversion.  
The solution\cite{dw}\cdash\cite{dolgov_hansen} of this equation in the
relevant range of parameters gives the following expression for the
cosmological density of relic sterile neutrinos:
\beq
\Omega_s \approx 0.3 \left ( \frac{\sin^2 \theta}{10^{-9}} \right )
\left ( \frac{m_s}{10\, {\rm keV} } \right )^2 
\eeq
The band of the masses and mixing angles consistent with dark matter
is shown in Fig.\ref{figure:range}.

Observations of the power spectrum of the Lyman-$\alpha$ forest clouds at
high redshift show a significant structure on small scales. This requires a
small collisionless damping scale associated with a dark matter particle,
in other words the dark matter must be sufficiently ``cold''. 
This and other constraints force the sterile neutrino mass to be greater
than 2.6~keV.\cite{Fuller,silk}  In Fig.\ref{figure:range}, the region
labeled ``too warm'' shows the boundary of the allowed range of masses. 

If the (unknown {\em a priori}) lepton asymmetry of the universe is
sufficiently large, then the sterile neutrinos can be produced through
resonant Mikheev-Smirnov-Wolfenstein\cite{msw} (MSW) oscillations in the
early universe.\cite{shi_fuller} These neutrinos are non-thermal and cold
because the adiabaticity condition selects the low-energy part of the
neutrino spectrum.

Most of the cosmological constraints can be evaded  if 
inflation ended with a low-temperature reheating.\cite{lowT}
In this case, the cosmological upper bound on the mixing angle is weaker,
and the allowed parameter space for the pulsar kicks extends to the lower
masses and the larger mixing angles.\cite{lowT}

\section{Pulsar kicks from active--sterile neutrino oscillations} 

We will now examine the range of parameters in which the emission of
sterile neutrinos from the core is sufficiently strong and anisotropic to
give the pulsar a kick.  There are three possible regimes for the sterile
neutrino emission.  Depending on the mass and the mixing angle, there may
or may not be a resonant conversion of the active to sterile neutrinos at
some density in a hot neutron star.  If there is an MSW resonance, the
position of the resonance point depends on the density and the magnetic
field.  The latter introduces the required anisotropy.  In the absence of
the MSW resonance, an off-resonance emission from the entire volume of the
neutron star core is possible.  We will see that this emission is efficient
only after the matter potential has evolved from its initial value to
nearly zero.  This important evolution\cite{Fuller} requires some time,
which, in turn imposes constraints on the masses and mixing angles.  We
will consider the following three possibilities for the pulsar kick:
\begin{itemize} 
\item MSW resonance in the core ($\rho > 10^{14}\, {\rm g/cm}^{3}$)
\item MSW resonance outside the core ($\rho < 10^{14}\, {\rm g/cm}^{3}$)
\item an off-resonance emission from the core 
\end{itemize} 
We will see that the three regimes are probably mutually exclusive.  For
example, for all the masses that are consistent with the resonance, the
matter potential evolves very slowly, and there is no significant emission
from the core off-resonance.

\subsection{MSW resonance in the core}
\label{sec_res_core}

Let us consider neutrino cooling during the first 10--15 seconds after the
formation of a hot proto-neutron star.  For simplicity we will assume that
it has a uniform (dipole) magnetic field $\vec{B}$.  Neutrino oscillations
in a magnetized medium are described by an effective potential that depends
on the magnetic field\cite{magn} in the following way:
\begin{eqnarray}
V(\nu_{\rm s}) & = & 0  \label{Vnus} \\
V(\nu_{\rm e})& = & -V(\bar{\nu}_{\rm e}) =  V_0 \: (3 \, Y_e-1+4 \,
Y_{\nu_{\rm e}}) \label{Vnue} \\ 
V(\nu_{\mu,\tau}) & = & -V(\bar{\nu}_{\mu,\tau}) = V_0 \: ( Y_e-1+2 \, 
Y_{\nu_{\rm e}}) \ 
+\frac{e G_{_F}}{\sqrt{2}} \left ( \frac{3 N_e}{\pi^4} 
\right )^{1/3}
\frac{\vec{k} \cdot \vec{B}}{|\vec{k}|} \label{Vnumu}
\end{eqnarray}
where $Y_e$ ($Y_{\nu_{\rm e}}$) is the ratio of the number density of
electrons (neutrinos) to that of neutrons, $\vec{B}$ is the magnetic field,
$\vec{k}$ is the neutrino momentum, $V_0=10 {\rm eV} \frac{\rho}{10^{14}
{\rm g/ cm}^{3} }$.  The magnetic field dependent term in equation
(\ref{Vnumu}) arises from polarization of electrons and {\em \sf not} from
a neutrino magnetic moment, which in the Standard Model is small and which
we will neglect. (A large neutrino magnetic moment can result in a pulsar
kick through a somewhat different mechanism,\cite{voloshin}  discussed
below.)

The condition for a resonant MSW conversion $\nu_i \leftrightarrow
\nu_j$ is

\begin{equation}
\frac{m_i^2}{2 k} \: \cos \, 2\theta_{ij} + V(\nu_i) = 
\frac{m_j^2}{2 k} \: \cos \, 2\theta_{ij} + V(\nu_j)  
\label{res}
\end{equation}
where $\nu_{i,j}$ can be either a neutrino or an anti-neutrino. 

In the presence of the magnetic field, the resonance condition (\ref{res})
for $\nu_{a}\rightarrow \nu_s$ ($a=\mu,\tau$) conversions is
satisfied at different distances $r$ from the center, depending on the
value of the $(\vec{k} \cdot \vec{B})$ term in (\ref{Vnumu}). The average
momentum carried away by the neutrinos depends on the temperature of the
region from which they escape.  The deeper inside the star, the higher is
the temperature during the neutrino cooling phase.  Therefore, neutrinos
coming out in different directions carry momenta which depend on the
relative orientation of $\vec{k}$ and $\vec{B}$.  This causes the asymmetry
in the momentum distribution.

\begin{figure}[ht]
\centerline{\epsfxsize=5in\epsfbox{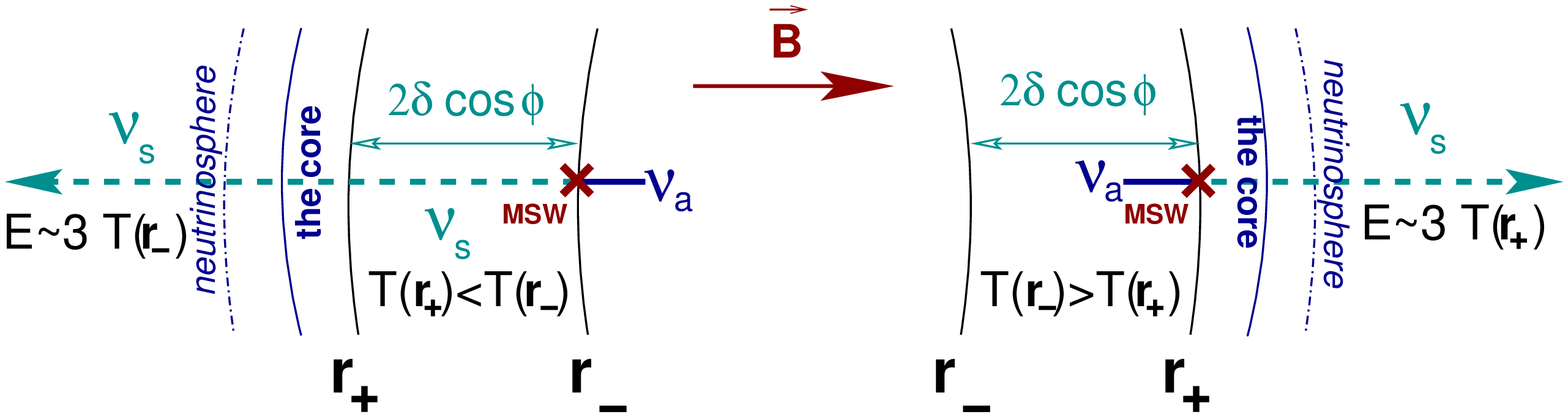}}   
\caption{ For MSW resonance {\em in the core}, the sterile neutrino energy
  depends on the temperature around the resonance point. 
}
\label{figure:core}
\end{figure}

The surface of the resonance points is 
\begin{equation}
r(\phi) = r_0 + \delta \: \cos \phi, 
\label{delta_r}
\end{equation}
where $\cos \, \phi= (\vec{k} \cdot \vec{B})/(k B)$ and $\delta$ is determined
by the equation 
$(d N_n(r)/dr) \delta \approx 
e \left ( 3 N_e/\pi^4 \right )^{1/3} B$.
This yields 

\begin{equation}
\delta = 
\frac{e \mu_e}{ \pi^2} \: B \left / \frac{dN_n(r)}{dr} \right. ,
\label{delta}
\end{equation}
where $\mu_e \approx (3 \pi^2 N_e)^{1/3} $ is the chemical potential of the
degenerate (relativistic)  electron gas.

In the core of the neutron star, at densities above $10^{14}$~g/cm$^3$, one
can assume the black-body radiation luminosity in sterile neutrinos: 
\beq
F_{\nu_s}({\bf r}) \propto T^4(r). 
\label{black_body}
\eeq 
Then the asymmetry in the momentum distribution is 
\begin{equation}
\frac{\Delta k_s}{k_s} \approx \frac{1}{3}\frac{T^4(r+\delta) -T^4(r-\delta)
}{T^4(r)} 
\approx 
\frac{4}{3} \frac{1}{T} \frac{dT}{dr} (2 \delta), 
\end{equation}
where a factor (1/3) represents the result of integrating 
over angles. 

Now we use the expression for $\delta$ from eq.~(\ref{delta}) and replace
the ratio of derivatives $(\frac{dT}{dr})/(\frac{dN_n}{dr})$ by
$\frac{dT}{dN_n}$: 
\beq 
\frac{\Delta k_s}{k_s} 
\approx \frac{2e}{3 \pi^2} \: \left (
\frac{\mu_e}{T} \frac{dT}{dN_n} \right) B.
\label{dk1}
\end{eqnarray}
To calculate the derivative in (\ref{dk1}), we assume approximate thermal
equilibrium.  Then one can use the relation between the
density and the temperature of a non-relativistic Fermi gas: 
\begin{equation}
N_n=\frac{2(m_n T)^{3/2}}{\sqrt{2} \pi^2}
\int \frac{\sqrt{z} dz}{e^{z-\mu_n/T}+1} 
\label{fermi}
\end{equation}
where $m_n$ and $\mu_n$ are the neutron mass and chemical potential. 
The derivative $(dT/dN_n)$ can be computed from (\ref{fermi}).  Finally,
\begin{equation}
\frac{\Delta k_s}{k_s} = 
\frac{8 e\sqrt{2}}{\pi^2} \: 
\frac{\mu_e \mu_n^{1/2}}{m_n^{3/2}T^2} \ B 
\end{equation}
We have assumed that only one of the neutrino species undergoes a resonance
transition into a sterile neutrino.  The energy, however, is shared between
6 species of active neutrinos and antineutrinos.  Therefore, the final
asymmetry due to anisotropic emission of sterile neutrinos is 6 times
smaller: 
\bea
\frac{\Delta k_s}{k} & = & \frac{1}{6}\frac{\Delta k_s}{k_s} =
\frac{4 e\sqrt{2}}{3 \pi^2} \: 
\frac{\mu_e \mu_n^{1/2}}{m_n^{3/2}T^2} \ B = \nonumber \\ & = & 
0.01 
\left ( \frac{\mu_e}{ 100 \rm \,MeV} \right )
\left ( \frac{\mu_n}{ 80 \rm \,MeV} \right )^{1/2}
\left ( \frac{20 \rm \,MeV}{T} \right )^2
\left ( \frac{B}{ 3\times 10^{16} {\rm \,G} }\right )
\label{dk2}
\eea
This estimate\cite{ks97} can be improved by considering a more detailed
model for the neutrino transport and by taking into account time evolution
of chemical potentials discussed below.  However, it is clear that the
magnetic field inside the neutron star should be of the order of
$10^{16}$~G.  The approximation used in equation (\ref{black_body}) holds
as long as the resonant transition occurs deep in the core, at density of
order $10^{14} \, {\rm g\,cm^{-3}}$.  This, in turn, means that the sterile
neutrino mass must be in the keV range.  We note that theoretical models of
neutrino masses can readily produce a sterile neutrino with a required
mass.\cite{farzan,babu}  The corresponding region of parameters is shown as
region ``1'' in Fig.~\ref{figure:range}.

\subsection{Resonance at densities below $ 10^{14}$g/cm$^3$}
\label{sec_res_lessdense}

For smaller masses, the resonance occurs at smaller densities.  Outside the
core, fewer neutrinos are produced, while there is a flux of
neutrinos diffusing out of the core.  Therefore, the approximation
(\ref{black_body}) is not valid.  

\begin{figure}[ht]
\centerline{\epsfxsize=5in\epsfbox{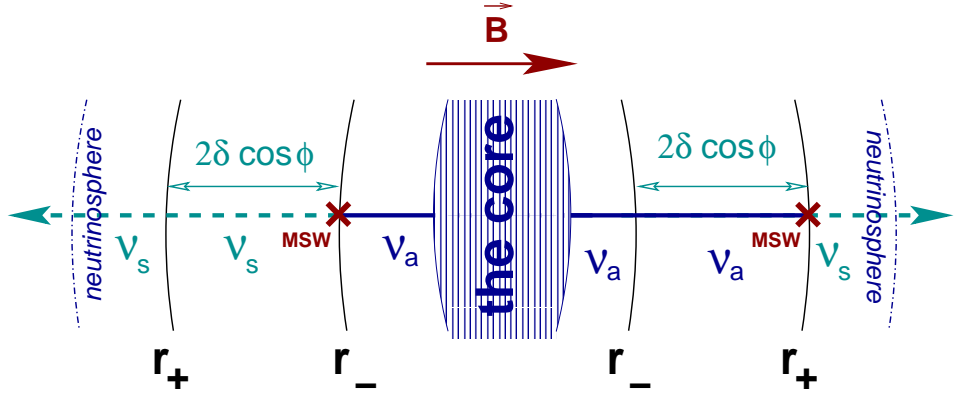}}   
\caption{ For MSW resonance {\em outside the core}, the neutrino passes
between $r_-=r_0-\delta \cos \phi$ and $r_+=r_0+\delta \cos \phi$ as a
sterile $\nu_s$ on one side of the star, while it still propagates as an 
active $\nu_a$ on the other side.  The active neutrinos interact and
deposit some extra momentum on the right-hand side, 
between $r_-$ and $r_+$.  Since the neutron star is a gravitationally bound
object, the momentum deposited asymmetrically in its outer layers gives the
whole star a kick. }
\label{figure:layer}
\end{figure}

Outside the core, the active neutrinos can interact
with matter and deposit momentum to the neutron star medium.  After an
active neutrino is converted into a sterile neutrino, it no longer
interacts with matter and comes out of the star.  
%
%
The cross section for active neutrino interactions in matter 
is $\sigma \sim \: G_{_F}^2 E_\nu^2 $, where
$E_\nu$ is the neutrino energy.  If the resonant conversion
$\nu_a\rightarrow \nu_s$ occurs at different depths for different
directions, the neutrinos may spend more time as active on one side of the
star than on the other side of the star, as shown in
Fig.~\ref{figure:layer}.  Hence, they deposit more momentum through their
interactions with matter on one side than on the other side.  Let us
estimate this difference.  (This argument was used before in application to
active neutrino transport.\cite{ks98}  Here we adopt it to sterile
neutrino.)

Depending on the magnetic field, the resonance lies at different depths,
eq.~(\ref{delta_r}).  Hence, the neutrinos on one side of the star pass an
extra layer of thickness $(2 \delta \cos\phi)$ as active, while the
neutrinos on the other side pass this layer as sterile.  The active
neutrinos going through a layer of nuclear matter with thickness $2\delta$
have an extra probability
\beq
P_\delta = (2\delta)\, \sigma \, N_n
\eeq
to interact and deposit momentum $k\sim E_\nu$ to the neutron star. This
momentum is not balanced by the neutrinos on the other side of the star
because they go through the this layer as sterile neutrinos.  

Obviously, the neutron star as a whole is a gravitationally bound object,
so any momentum deposited on one side of the star gives the whole neutron
star a kick.  

The difference in the momentum deposition per active neutrino between
the directions $\phi$ and $-\phi$ is
%
\beq
\frac{\Delta k_s}{k}  \sim  (2 \delta \cos \phi) \: 
N_n \: \sigma 
\sim \: G_{_F}^2 E_\nu^2 \: \frac{\mu_e}{Y_e} \: \frac{eB}{\pi^2} 
\: h_{N_e} \:  \cos \phi,
\eeq
where we have used eq.~(\ref{delta}) and introduced the scale height 
of the electron density $h_{N_e}=[d(\ln N_e)/dr]^{-1}$.

We take $Y_e\approx 0.1$, $E_\nu \approx 3 T\approx 10 $~MeV, $\mu_e\approx
50$~MeV, and $h_{N_e}\approx 6$~km.  We assume $T\approx 3$~MeV because it
is a realistic average temperature around the neutrinosphere, in agreement
with theoretical models as well as observations of the supernova
SN1987A.\cite{T_1987A} (This temperature is lower than the core
temperature used earlier.)

After integrating over angles and taking into account that only one
neutrino species undergoes the conversion, we obtain the final result for the
asymmetry in the momentum deposited by the neutrinos.  

\begin{equation}
\frac{\Delta k_s}{k} = 0.03 
\left ( \frac{T}{3\, \rm MeV} \right )^2
\left ( \frac{\mu}{50 \, \rm MeV} \right )
\left ( \frac{h}{6\, \rm km} \right )
\left ( \frac{B}{10^{15} \, {\rm G}} \right ),
\label{final} 
\end{equation}
This is, clearly, a sufficient asymmetry for the pulsar kick. The
corresponding region of parameters is shown as region ``2'' in
Fig.~\ref{figure:range}.
 
The above estimates are valid as long as $\delta$ is much smaller than the
mean free path.  One can also describe the propagation of neutrinos in this
region using the so called diffusion approximation.\cite{landau}  It was
used for the neutrino transport by Schinder and Shapiro\cite{shapiro} in
planar approximation and was applied to the pulsar kicks by Barkovich~{\em
et al.}.\cite{barkovich,barkovich_s}

\subsection{Off-resonance transitions} 
\label{sec_offres}

Let us now consider the case of the off-resonance emission from the core.
This possibility was discussed qualitatively in section~\ref{sec_why}.  Now
we want to determine the neutrino parameters consistent with the kick
mechanism. 

For masses of a few keV, the resonant condition is not satisfied
anywhere in the core.  In this case, however, the off-resonant
production of sterile neutrinos in the core can occur through ordinary urca
processes.  A weak-eigenstate neutrino has a $\sin^2\theta $ admixture
of a heavy mass eigenstate $\nu_2$.  Hence, these heavy neutrinos can be
produced in weak processes with a cross section suppressed by $\sin^2\theta
$. 

Of course, the mixing angle in matter $\theta_m$ is not the same as it is
in vacuum, and initially $\sin^2\theta_m \ll \sin^2\theta$.  However, as
Abazajian, Fuller, and Patel\cite{Fuller} have pointed out, in the presence
of sterile neutrinos the mixing angle in matter quickly evolves toward its
vacuum value.  When $\sin^2\theta_m \approx \sin^2\theta$, the production
of sterile neutrinos is no longer suppressed, and they can take a fraction
of energy out of a neutron star.  We note in passing that time evolution of
the matter potential may be important for a number of other
reasons.\cite{changeinV}

Following Abazajian, Fuller, and Patel,\cite{Fuller} one can estimate the
time it takes for the matter potential to evolve to zero from its initial
value $V^{(0)}(\nu_e)\simeq (-0.2 ...+ 0.5) V_0$.  The time scale for this
change to occur through neutrino oscillations off-resonance is    
\begin{eqnarray}
\label{timeeqoffres}
 \tau_{_V}^\mathrm{off-res}  & \simeq &
\frac{4 \sqrt{2} \pi^2 m_n}{G_{\!\!_F}^3 \rho}
\frac{ (V^{(0)}(\nu_e))^3}{(\Delta m^2)^2 \sin^2 2 \theta } \frac{1}{\mu^3}
\\
& \sim & 
 \frac{6 \times 10^{-9} s}{\sin^2 2 \theta} 
\left (\frac{V^{(0)}(\nu_e)}{0.1 \mathrm{eV}} \right )^3 
\left (\frac{50 \mathrm{MeV}}{\mu} \right )^3 \left ( \frac{ 
10 \mathrm{keV}^2
}{\Delta m^2
} \right )^2. \nonumber   
\label{timeeqoffresnumerical}
\end{eqnarray}

As long as this time is much smaller than 10 seconds, the mixing angle in
matter approaches its vacuum value in time for the sterile neutrinos to
take out some fraction of energy from a cooling neutron star.  

The urca processes produce ordinary neutrinos with some asymmetry depending
on the magnetic field.\cite{drt}  The same asymmetry is present in the
production cross sections of sterile neutrinos.  However, unlike the active
neutrinos, sterile neutrinos escape from the star without rescattering.
Therefore, the asymmetry in their emission is not washed out as it is in
the case of the active neutrinos.\cite{eq}  Instead, the asymmetry in
emission is equal the asymmetry in production.
 
\begin{figure}[ht]
\centerline{\epsfxsize=4.1in\epsfbox{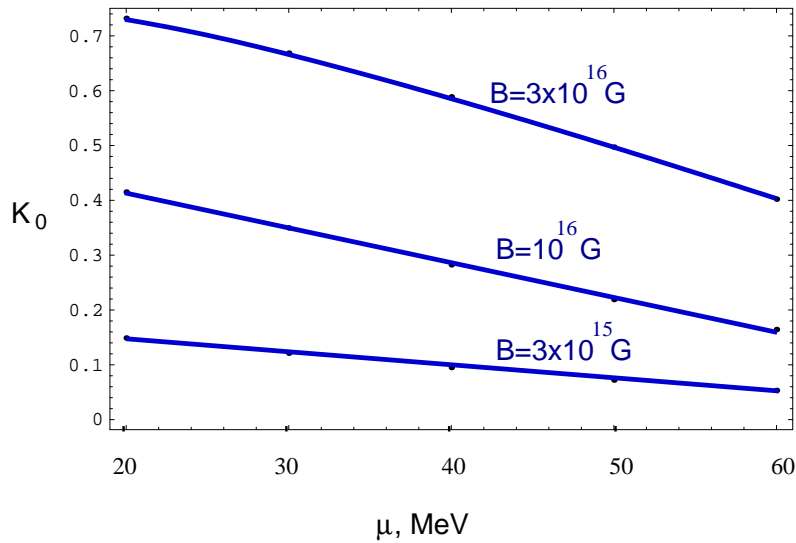}}   
\caption{The fraction of electrons in the lowest Landau level as a function
chemical potential.  
The value of the magnetic field is shown next to each curve.  }
\label{figure:asymmT20}
\end{figure}

The number
of neutrinos $dN$ emitted into a solid angle $d\Omega $ can be written as
\begin{equation}
\frac{dN}{d\Omega}= N_0(1+ \epsilon \cos \Theta_\nu ), 
\end{equation}
where $\Theta_\nu$ is the angle between the direction of the magnetic field
and the neutrino momentum, and $N_0$ is some normalization factor.  The
asymmetry parameter $\epsilon$ is equal
\begin{equation}
\epsilon = \frac{g_{_V}^2-g_{_A}^2}{g_{_V}^2+3g_{_A}^2} 
k_0   
\left ( \frac{  E_{\rm s}}{ E_{\rm tot}} \right ),  
\end{equation} 
where $g_{_V}$ and $g_{_A}$ are the axial and vector couplings, ${ E_{\rm
tot}}$ and $ { E_{\rm s}} $ are the total neutrino energy and the energy
emitted in sterile neutrinos, respectively.  The number of electrons in the
lowest Landau level, $k_0$, depends on the magnetic field and the chemical
potential $\mu$ as shown in Fig.~\ref{figure:asymmT20}.

The momentum asymmetry in the neutrino emission is 
\begin{equation}
\epsilon \sim 0.02 
\left ( \frac{k_0}{0.3}\right ) 
\left ( \frac{ r_{_E}}{0.5} \right), 
\label{epsilon_final} 
\end{equation}  
where $r_{_E}$ is the fraction of energy carried by the sterile neutrinos.
To satisfy the constraint based on the observation of neutrinos from
supernova SN1987A, we require that $r_{_E}<0.7$.  As can be seen from
Fig.~\ref{figure:asymmT20}, the asymmetry in equation (\ref{epsilon_final})
can be of the order of a few per cent, as required, for magnetic fields
$10^{15}-10^{16}$~G.

Surface magnetic fields of ordinary radio pulsars are estimated to be of
the order of $10^{12}-10^{13}$G.  However, the magnetic field inside a
neutron star may be much higher,\cite{magnetic,dt,magnetars} probably up to
$10^{16}$G.  The existence of such a strong magnetic field is suggested by
the dynamics of formation of the neutron stars, as well as by the stability
of the poloidal magnetic field outside the pulsar.\cite{dt} Moreover, the
discovery of soft gamma repeaters and their identification as
magnetars,\cite{magnetars} {\em i.e.}, neutron stars with {\em surface}
magnetic fields as large as $10^{15}$~G, gives one a strong reason to
believe that the interiors of many neutron stars may have magnetic fields
as large as $10^{15}-10^{16}$~G.  There are also plausible
physical mechanisms that can generate such a large magnetic field
inside a cooling neutron star.\cite{magnetic,dt,zeldovich}

\subsection{Pulsar kicks from the active neutrinos alone?} 

One can ask whether the sterile neutrino is necessary and whether the
oscillations of active neutrinos alone could explain the pulsar kicks.  The
interactions of muon and tau neutrinos in nuclear matter are characterized
by a smaller cross section than those of the electron neutrinos.  This is
because the electron neutrinos $\nu_e$ interact through both charged and
neutral currents with electrons, while $\nu_\mu$ and $\nu_\tau$ interact
with electrons 
through neutral currents alone.  Therefore, nuclear matter is more
transparent to $\nu_e$ than to $\nu_{\mu,\tau}$, $\bar{\nu}_{\mu,\tau}$.
As a result, the surface of last scattering for $\nu_{\mu,\tau}$ and
$\bar{\nu}_{\mu,\tau}$ lies (about a kilometer) deeper than that of
$\nu_e$.  The electron antineutrino can interact through charged currents
with {\em positrons} while they are present in nuclear matter.  The
$\bar{\nu}_e$ mean free path starts out closer to that of $\nu_e$, but, as
the number of positrons diminishes during the cooling period, this mean
free path increases and becomes comparable to that of $\nu_{\mu,\tau}$ and
$\bar{\nu}_{\mu,\tau}$.

Since the $\mu-$ and $\tau-$neutrinospheres lie inside the electron
neutrinosphere, it is possible that neutrino oscillations could convert a
$\nu_e$ into $\nu_\mu$ or $\nu_\tau$ at some point between the two
neutrinospheres, where the $\nu_e$ is {\em trapped}, but the
$\nu_{\mu,\tau}$ is free-streaming.  Then the shift in position of the MSW
resonance would result in an anisotropy of the outgoing momentum.  This
mechanism could explain the pulsar kicks, but it would require one of the
active neutrino masses to be of order $100$eV.\cite{ks96}  This is not
consistent with the present data.

\subsection{What if neutrinos have a large magnetic moment?} 

The neutrino magnetic moment in the Standard Model is very small, $\mu_\nu
\approx 3\times 10^{-19}(m_\nu/{\rm eV})\mu_{_B} $, where $\mu_{_B}
=e/2m_e$ and $m_\nu$ is the (Dirac) neutrino mass.  This is why we have so
far neglected any effects of direct neutrino interactions with the magnetic
field.  

However, the present experimental bounds allow the neutrino magnetic
moments to be as large as $10^{-12}\mu_{_B} $.  If, due to some new
physics, the neutrino magnetic moment is large, it may open new
possibilities for the pulsar kick.  Voloshin\cite{voloshin} has proposed an
explanation of the pulsar kick based on the resonant spin-flip and
conversion of the left-handed neutrinos into the right-handed neutrinos,
which then come out of the neutron star.  Voloshin argued that the magnetic
field inside a neutron star may be irregular and may have some
asymmetrically distributed ``windows'', through which the right-handed
neutrinos could escape.  The resulting asymmetry may, indeed, explain the 
pulsar velocities.

Other kinds of new physics may cause the pulsar kicks as well\cite{ext}.  

\subsection{Spin-kick from neutrinos} 

Spruit and Phinney\cite{phinney} have argued that the pulsar rotational
velocities may also be explained by the kick received by the neutron stars
at birth.  The core of the progenitor star is likely to co-rotate with the
whole star until about 10 year before the collapse.  This is because the
core should be tied to the rest of the star by the magnetic field.
However, then the angular momentum of the core at the time of collapse is
$10^3$ times smaller than the angular momentum of a typical pulsar.  Spruit
and Phinney have pointed out that the kick that accelerates the pulsar can
also spin it up, unless the kick force is exerted exactly head-on.

The neutrino kick can be strongly off-centered, depending on the
configuration of the magnetic field.  If the magnetic field of a pulsar is
offset from the center, so will be the force exerted on the pulsar by the
anisotropic emission of neutrinos.  This mechanism may explain
simultaneously the high spatial velocities and the unusually high rotation
speeds of nascent neutron stars.\cite{phinney}  E.S.~Phinney 
has suggested\cite{phinney_private} that a highly off-centered magnetic
field could be generated by a {\it thin-shell} dynamo in a hot
neutron star.  Since the neutron star is cooled from the outside, a
convective zone forms near the surface and, at later times, extends to the
interior.  While convection takes place in the spherical shell, the dynamo
effect can cause a growth in the magnetic field.
Thin-shell dynamos are believed to be responsible for generating magnetic
fields of Uranus and Neptune,\cite{rs,un} which, according to the {\em
Voyager~2} measurements,\cite{voyager} are both off-centered and tilted
with respect to the axis of rotation.  Unlike other planets, which have
convection in the deep interior and end up with a well-centered dipole 
field, Uranus and Neptune have thin spherical convective zones near the
surface, which explains the peculiarity of their dynamos.  During the
first few seconds after the supernova collapse, convection in a neutron
star also takes place in a spherical layer near the surface.  The
thin-shell dynamo can, in principle, generate an off-center magnetic field
in a neutron star, just like it does in Uranus and
Neptune.\cite{phinney_private}

\section{Observational consequences and experimental searches}

The mixing angle $\theta$, consistent with the neutrino kick and also with
dark matter, is very small: $\sin\theta \sim 10^{-5}-10^{-4}$.  Hence, it
is unlikely that the existence of the sterile neutrino with the required
mixing can be tested in a laboratory experiment.  However, some
astrophysical observations can be used to verify or rule out our scenario.

\subsection{Search for the relic sterile neutrinos with mass 2--20~keV.} 

The parameter space allowed for the pulsar kicks\cite{fkmp} overlaps nicely
with that of dark-matter sterile neutrinos.\cite{Fuller,dw}  Relic sterile
neutrinos in this range should make up the galactic halos.  For smaller
mixing angles, some part of dark matter is sterile neutrinos.  In any case,
there should exist some population of sterile neutrinos left over after the
Big Bang.

\begin{figure}[ht]
\centerline{\epsfxsize=5in\epsfbox{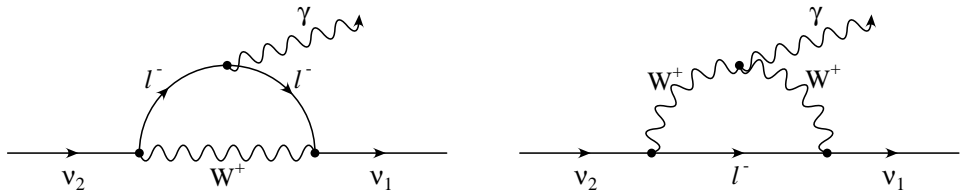}}   
\caption{
Radiative decay of sterile neutrinos, $\nu_2\rightarrow \nu_1 \gamma$.  The
X-rays produced by these decays can be detected by the X-ray telescopes,
such as {\em Chandra}, {\em XMM-Newton}, and the future {\em
  Constellation-X}.  
}
\label{figure:nusdecay}
\end{figure}

The relic sterile neutrinos with mass in the 2--20~keV range can decay into
three lighter neutrinos, or into a lighter neutrino and a photon.  The
Feynman diagrams for the latter channel of decay are shown in
Fig.\ref{figure:nusdecay}.  The rate of the radiative decay is
\begin{equation}
\Gamma_\gamma \approx 6.8 \times 10^{-33}\,{\rm
s^{-1}}\ \left(\frac{\sin^2 2\theta}{10^{-10}}\right)
\left(\frac{m_s}{1\,\rm keV}\right)^5. 
\end{equation}
Although $\tau = \Gamma^{-1}\sim 10^{25}-10^{33}\,$s is much longer than
the age of the universe, there are, nevertheless, enough decays in the
clusters of galaxies for the photons to be observed.\cite{aft} Since
$\nu_2\rightarrow \nu_1 \gamma$ is a two-body decay, the photon energy is
equal $(m_s/2)$, which is in the 1--10 keV range for the masses of interest
to us.  These photons should be detectable by the X-ray telescopes.  {\em
Chandra} and {\em XMM-Newton} can exclude part of the parameter
space\cite{aft} shows in Fig.\ref{figure:range}.  The future {\em
Constellation-X} can probably explore the entire allowed range of
parameters.

To detect the relic sterile neutrinos, one should look for an isolated line
that cannot be identified with another source, such as interstellar gas.
One can try to distinguish the lines in the gas emission from the one due
to dark matter by comparing the strengths of signals from regions with
different temperatures.  The detection strategy is discussed in detail by
Abazajian, Fuller, and Tucker.\cite{aft}

\subsection{Gravity waves from a pulsar kick due to neutrinos} 

In the event of a nearby supernova, the neutrino kick can produce gravity
waves that could be detected by LIGO and LISA.\cite{loveridge,cuesta}
These gravity waves can be produced in several ways.  

Obviously, the departure from spherical symmetry is a necessary condition
for generating the gravity waves.  A neutron star being accelerated by
neutrinos is not moving fast enough to generate gravitational waves from it
own motion.  However, the anisotropy in the outgoing neutrinos turns out to
be sufficient to produce an observable signal in the event of a nearby
supernova. 

\begin{figure}[ht]
\centerline{\epsfxsize=5.2in\epsfbox{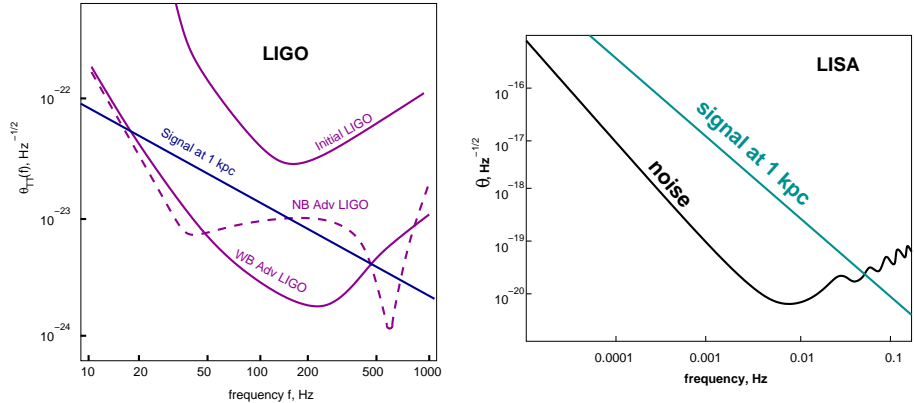}}   
\caption{
Gravity waves signal at LIGO and LISA calculated by 
L.C.~Loveridge.
}
\label{figure:grav_waves}
\end{figure}

Most of the neutrinos come out isotropically and can be neglected.
However, a few per cent of asymmetrically emitted neutrinos move along the
direction of the magnetic field.  In general, the magnetic field is not
aligned with the axis of rotation, and, therefore, the outgoing neutrinos
create a non-isotropic source for the waves of gravity. (Water jet produced
by a revolving lawn sprinkler is probably a good analogy for the geometry
of this source.)  The  signal was calculated by
L.~Loveridge.\cite{loveridge} It can be observed by advanced LIGO or LISA
if a supernova occurs nearby, as shown in Fig.~\ref{figure:grav_waves}.
Alternatively, the neutrino conversion itself may cause gravity waves
coming out of the core.\cite{cuesta}


\subsection{{$B-v$} correlation?} 

Unfortunately, the neutrino kick mechanism does not predict a correlation
between the direction of the surface magnetic fields and the pulsar
velocity.  The kick velocity is determined by the magnetic field {\em
inside} the hot neutron star during the {\em first seconds} after the
supernova collapse.  Astronomical observations can be used to infer the
{\em surface} magnetic fields of pulsars some {\em millions of years}
later.  The relation between the two is highly non-trivial because of the
complex evolution the magnetic field undergoes in a cooling 
neutron star.  Let us outline some stages of this evolution.

Immediately after the formation of the hot neutron star the magnetic field
is expected to grow due to differential rotation, thermal
effects,\cite{magnetic} and convection.\cite{zeldovich} The dynamo effect
can probably account for the growth of the magnetic field to about
$10^{15}-10^{16}$~G.\cite{dt} 

The growth of the magnetic field takes place during the first ten seconds
after the supernova collapse, in part because the neutrino cooling causes
convection.  At the same time, during the neutrino cooling phase, the
neutron star receives a kick.  The magnetic field relevant for the kick is
the average interior magnetic field during the first 10 seconds.  There is
no reason to believe that it has the same direction or magnitude as the
surface field at the end of the neutrino cooling phase.  This, however, is
only one of several stages in the evolution of the magnetic field.  Next,
at some temperature below $0.5$~MeV, the nuclear matter becomes a type-II
superconductor.  The magnetic field lines form flux tubes, reconnect, and
migrate.  Next, over millions of years, the pulsar rotation converts the
magnetic field energy into radio waves and causes the field to evolve even
further.  The end result of this evolution is, of course, a configuration
of magnetic fields that is very different from what it was five seconds
after the onset of the supernova.

Clearly, the magnetic field inside a hot young neutron star is not expected
to have much correlation with the surface field of a present-day pulsar.
Some naive analyses of the $B-v$ correlation have ignored the magnetic
field evolution and have reached incorrect conclusions.\cite{birkel_toldra}

\section{Conclusion} 

An asymmetric neutrino emission from a cooling neutron star can explain the
observed pulsar velocities.  The asymmetry may be caused by neutrino
oscillations in the magnetized nuclear matter if there is a sterile
neutrino with mass in the 2--20~keV range and a small mixing with ordinary
neutrinos.  It is intriguing that the same particle is a viable dark matter
candidate.  We know that at least one particle beyond the Standard Model
must exist to account for dark matter.  This particle may come as part of a
``package'', for example, if supersymmetry is realized in nature.  However,
it may be that the dark matter particle is simply an
SU(3)$\times$SU(2)$\times$U(1)-singlet fermion, which has a small mixing
with neutrinos.

Future observations of X-ray telescopes may be able to discover the relic
sterile neutrinos by detecting keV photons from the sterile neutrino decay
in clusters of galaxies.  Finally, if gravitational waves are detected from
a nearby supernova, the signal may show the signs of a neutron star being
accelerated by an asymmetric neutrino emission.

\section*{Acknowledgments}

The author thanks J.C.~D'Olivo and E.S.~Phinney for helpful discussions.  
This work was supported in part by the U.S. Department of Energy grant
DE-FG03-91ER40662 and the NASA Astrphysics Theory Program grant NAG5-13399.


\end{document}